\documentclass{Interspeech}

\interspeechcameraready

\title{Accelerating Autoregressive Speech Synthesis Inference With Speech Speculative Decoding}

\author[affiliation={1}]{Zijian}{Lin}
\author[affiliation={2}]{Yang}{Zhang}
\author[affiliation={2}]{Yougen}{Yuan}
\author[affiliation={2}]{Yuming}{Yan}
\author[affiliation={1}]{Jinjiang}{Liu} 
\author[affiliation={1,\ast}]{Zhiyong}{Wu}
\author[affiliation={2}]{Pengfei}{Hu}
\author[affiliation={2}]{Qun}{Yu}

\affiliation{Shenzhen International Graduate School}{Tsinghua University}{China}
\affiliation{Platform and Content Group}{Tencent}{China}
\email{linzj24@mails.tsinghua.edu.cn, autozhang@tencent.com, zywu@sz.tsinghua.edu.cn}

\usepackage{comment}
\usepackage[ruled]{algorithm2e}
\usepackage{graphicx} 
\usepackage{subcaption} 
\usepackage{float} 
\begin{document}

\maketitle

\renewcommand{\thefootnote}{\fnsymbol{footnote}}
\footnotetext[1]{Corresponding author.}
\renewcommand{\thefootnote}{\arabic{footnote}}

\begin{abstract}
Modern autoregressive speech synthesis models leveraging language models have demonstrated remarkable performance. However, the sequential nature of next token prediction in these models leads to significant latency, hindering their deployment in scenarios where inference speed is critical. In this work, we propose Speech Speculative Decoding (SSD), a novel framework for autoregressive speech synthesis acceleration. Specifically, our method employs a lightweight draft model to generate candidate token sequences, which are subsequently verified in parallel by the target model using the proposed SSD framework. Experimental results demonstrate that SSD achieves a significant speedup of 1.4x compared with conventional autoregressive decoding, while maintaining high fidelity and naturalness. Subjective evaluations further validate the effectiveness of SSD in preserving the perceptual quality of the target model while accelerating inference. \footnote{Speech samples: https://thuhcsi.github.io/interspeech2025-SSD/}

\end{abstract}

\section{Introduction}
\begin{figure}[h]
    \centering
    \begin{subfigure}[b]{0.48\linewidth}
        \centering
        \includegraphics[width=\textwidth]{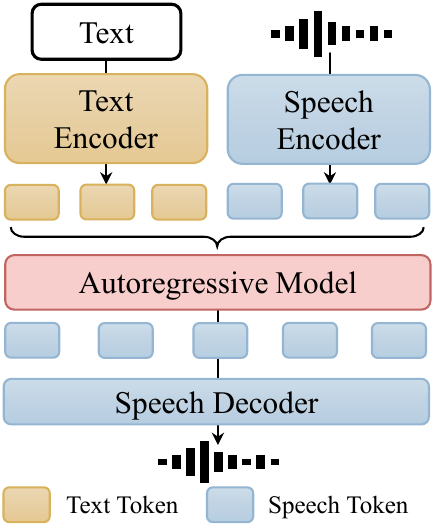}
        \caption{Overview of the autoregressive speech synthesis system.}
        \label{fig:sub1}
    \end{subfigure}
    \hfill
    \begin{subfigure}[b]{0.48\linewidth}
        \centering
        \includegraphics[width=\textwidth]{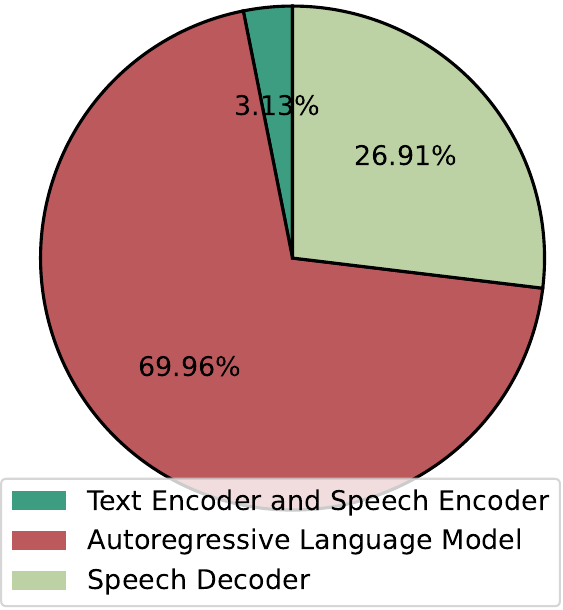}
        \caption{Inference time composition of CosyVoice 2. The speech decoder of CosyVoice 2 includes flow-matching model and vocoder.}
        \label{fig:sub2}
    \end{subfigure}
    \caption{Visualization of autoregressive speech synthesis system and inference time composition of CosyVoice 2.}
    \label{fig:combined}
    \vspace{-20pt}
\end{figure}

Modern autoregressive speech synthesis systems \cite{wang2023neural, zhang2023speak, anastassiou2024seed, du2024cosyvoice, du2024cosyvoice2, liao2024fish, guo2024fireredtts, kharitonov2023speak, lajszczak2024base,peng-etal-2024-voicecraft,yang2023uniaudio} often employ four core components to achieve human-like speech generation, as shown in \autoref{fig:sub1}. The text encoder tokenizes input text into fine-grained sequences, and in some cases, further transforms them into high-dimensional semantic representations. The speech encoder discretizes continuous audio into semantic speech token sequences through vector quantization. The autoregressive language model (LM) then processes these tokens, capturing long-range temporal dependencies and contextual patterns to predict sequential subsequent acoustic units. Finally, the speech decoder reconstructs these discrete speech tokens into high-fidelity waveform signals. This architecture enables precise prosodic control and exceptional coherence in synthesized speech, setting new benchmarks for naturalness through the powerful contextual modeling capabilities of the language model.

Despite the great success, Autoregressive (AR) models generate speech strictly sequentially, leading to computationally intensive and time-consuming inference. This inefficiency is caused by two key factors: first, speech token sequences are inherently long, as each token typically represents a short segment of speech (e.g., 25 ms or 50 ms) \cite{hsu2021hubert, 9625818, defossezhigh}, resulting in significantly longer sequences compared to text tokens; and second, the increasing parameter size of neural networks, driven by the pursuit of higher model performance, further amplifies computational demands. The inference time scales quadratically with sequence length \cite{vaswani2017attention}, while the computational complexity of the Transformer backbone grows quadratically, further exacerbating inefficiency. For instance, in CosyVoice 2 \cite{du2024cosyvoice2}, a state-of-the-art LM-based speech synthesis model, the autoregressive inference process of the LM takes up approximately 70\% of the total synthesis time, as illustrated in \autoref{fig:sub2}. This bottleneck has made inference acceleration a critical research focus in AR speech synthesis, driving exploration of non-autoregressive (NAR) alternatives \cite{kim2021conditional, kong23_interspeech} such as diffusion models \cite{chen-etal-2024-f5tts,JuWS0XYLLST000024} or flow matching \cite{eskimez2024e2, le2024voicebox}, which enable parallel generation and significantly improve inference speed.

\begin{figure*}[t]
  \centering
  \includegraphics[width=0.65\linewidth]{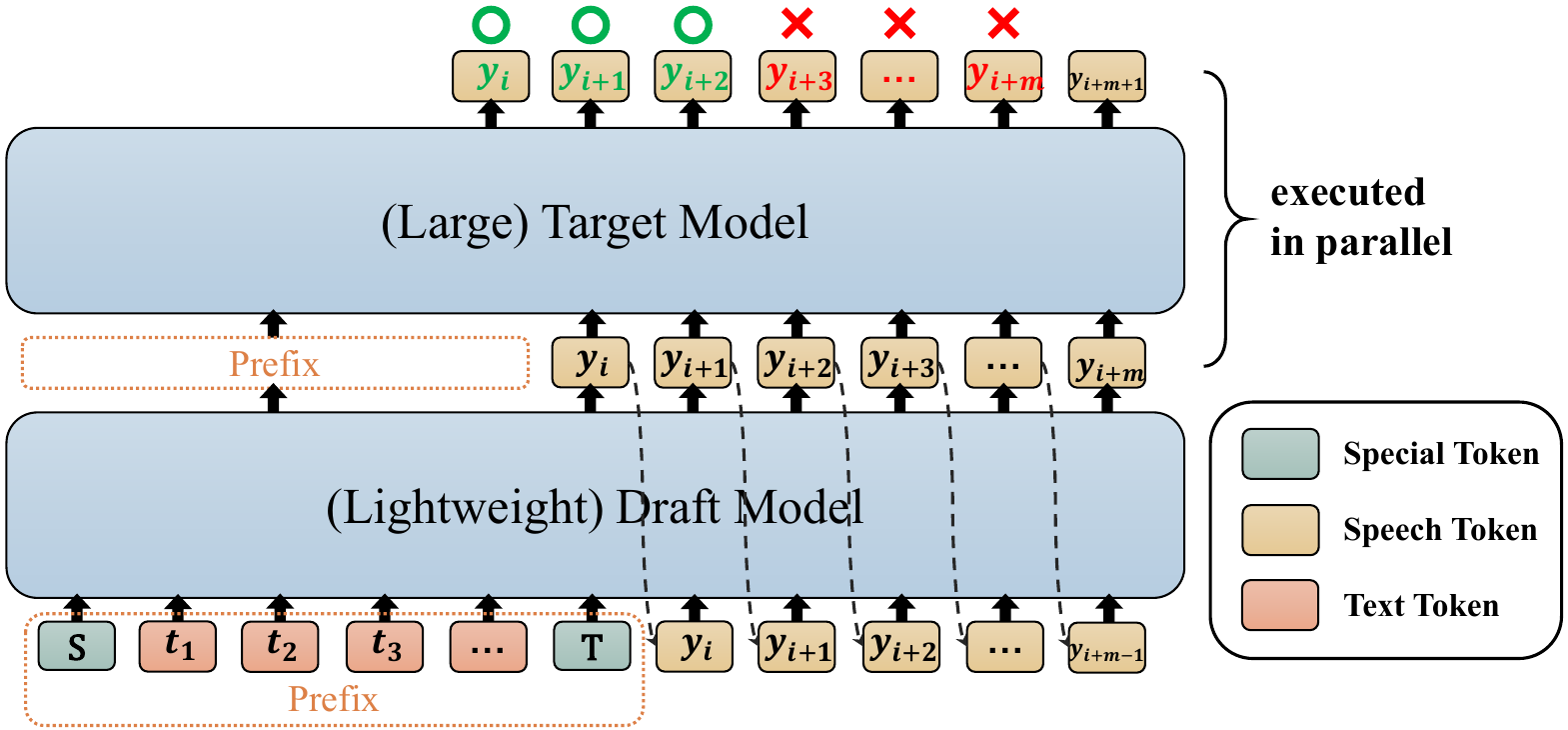}
  \vspace{-9pt}
  \caption{The Inference Architecture of proposed \textbf{Speech Speculative Decoding (SSD)} framework. The dashed line depicts autoregressive decoding in the draft model, followed by parallel verification by the target model. The prefix comprises all prompt tokens and previously generated tokens during inference. Green indicates accepted tokens, while red represents rejected tokens.}
  \label{fig:overall_architecture}
  \vspace{-19pt}
\end{figure*}

Speculative decoding \cite{chen2023accelerating, leviathan2023fast} accelerates text-based large language models (LLMs) by employing a small draft model to autoregressively generate candidate token sequences, which are then verified in parallel by a larger target model. Inspired by these approaches, there have been notable attempts \cite{10887855,10888194} to apply speculative decoding to accelerate autoregressive speech synthesis. However, they integrate multiple draft heads directly into the target model and tend to require fine-tuning the target model itself to enhance draft head quality. In contrast, we introduce Speech Speculative Decoding (SSD) with an independent draft architecture that maintains high-quality predictions without requiring target model fine-tuning. Moreover, unlike text tokens, which often have clear one-to-one mappings with words or phonemes, speech tokens exhibit diverse and variable mappings, as audio segments that sound nearly identical can have different token-level representations. This variability stems from the continuous and dynamic nature of speech signals. Traditional speculative decoding \cite{chen2023accelerating, leviathan2023fast}, designed for text, may result in suboptimal efficiency for speech synthesis, as their acceptance criteria fail to account for these nuanced relationships. To address this, we propose an improved acceptance criterion tailored to speech tokens, enabling more efficient verification of candidate sequences generated by the draft model. The adaptation in SSD finally achieves a significant speedup of 1.4$\times$ compared to conventional autoregressive decoding, while maintaining high fidelity and naturalness.


Our contributions can be summarized as follows. First, we present a novel SSD method targeting speech synthesis acceleration through two key mechanisms: a draft model to generate candidate token sequence, and a target model for parallel verification.
Second, we design an efficient way to construct the lightweight draft model for alignment.
Third, a tolerance factor is used to influence the acceptance rate of generated tokens, thus balancing the speech synthesis effect and inference speed.

\section{Methodology}

In this section, we demonstrate the proposed SSD acceleration framework, which comprises two core components: a target model (the large model requiring acceleration) and a draft model (a lightweight auxiliary model). 
As illustrated in \autoref{fig:overall_architecture}, the process unfolds in three phases: 1) Autoregressively generating 
$m$ candidate tokens using the draft model, 2) Parallel verification of these tokens in the target model via SSD, 3) Applying a speculative decoding-based acceptance strategy to validate the token sequences.

\subsection{Autoregressive Language Model for Speech Synthesis}

In an autoregressive speech synthesis system, a language model is applied to predict discrete tokens of speech. 
Our approach builds upon the open-source CosyVoice framework \cite{du2024cosyvoice,du2024cosyvoice2} for autoregressive speech synthesis, which uses language model to predict single stream of discrete speech tokens and then uses flow-matching model as the speech decoder.

Given a text-speech pair \(\{T, A\}\), where \(T\) denotes text sequence and \(A\) represents the speech signal, text \(T\) is tokenized into a more fine-grained sequence \(T = \{t_1, t_2, \ldots, t_N\}\) 
of length \(N\) via some methods such as grapheme-to-phoneme (g2p) transformation or BPE-based text tokenizer, 
and the speech encoder discretizes the continuous speech signal \(A\) into a sequence of discrete tokens \(Y = \{y_1, y_2, \ldots y_L\}\) of length \(L\). Then the language model predicts \(Y\) autoregressively taken text \(T\) as input. To input all the information into the language model, text tokens and speech tokens are typically concatenated into a single sequence via some special tokens and then uniformly fed into the model.

Therefore, in the setting of the zero-shot TTS, the language model \(\theta_{LM}\) learns to model the distribution of the target speech token sequence \(Y_{tgt}\):
\begin{equation}
\begin{split}
P(Y_{tgt}|&T_{tgt}, Y_{pt}, T_{pt};\theta_{LM}) = \\
&\textstyle\prod_{t = 1}^{L} P(Y_{tgt}^{t}|Y_{tgt}^{<t}, T_{tgt}, Y_{pt}, T_{pt};\theta_{LM}),
\end{split}
\end{equation}
where \(Y_{pt}\) denotes the audio prompt, \(T_{pt}\) represents the transcription of the audio prompt, and \(T_{tgt}\) means the target text. Traditional autoregressive models suffer from speed limitations due to their token-by-token serial generation process. This sequential dependency significantly reduces computational efficiency and brings to high latency.

\vspace{-10pt} 
\begin{algorithm}[h]
\caption{SSD for Speech Synthesis}
\KwIn{Prompt \(T_{in}\), target model \(M_{q}\), draft model \(M_{p}\), draft length \(L_d\), target length \(L_t\), tolerance factor \(\beta\)}
\KwOut{Generated speech signal}
Initialize \(L_{acc} \leftarrow 0\) \\
\While{\(L_{acc} < L_t\)}{
    \(p^d_1, \ldots, p^d_{L_d} \leftarrow M_{p}(T_{in})\) \\
    \(x_1, \ldots, x_{L_d} \sim p^d_1, \ldots, p^d_{L_d}\) \\
    \(q^s_1, \ldots, q^s_{L_d + 1} \leftarrow M_{q}(T_{in}, x_1, \ldots, x_{L_d})\) \\
    \For{\(i = 1\) \KwTo \(L_d\)}{
        Sample \(r_i \sim U[0,1]\) \\
        \eIf{\(r_i < \min(1, q^s_i / p^d_i) + \beta\)}{
            Append \(x_i\) to \(T_{in}\)
        }{
            Resample \(x'_i \sim \text{normlize}(\max(0, q^s_i - p^d_i))\) \\
            Append \(x'_i\) to \(T_{in}\) \\
            \textbf{break}
        }
    }
    \If{All tokens accepted}{
        Append \(x_{L_d+1} \sim q^s_{L_d+1}\) to \(T_{in}\)
    }
    Update \(L_{acc}\) based on accepted tokens
}
\label{alg:speculative}
\end{algorithm}

\vspace{-20pt} 
\subsection{Speech Speculative Decoding}

As shown in Algorithm \ref{alg:speculative}, 
speculative decoding accelerates autoregressive generation by leveraging a draft model \(M_p\) to propose candidate tokens and a target model \(M_q\) to verify them. In speech synthesis, we adapt this framework to address the inherent ambiguity of speech tokens, which lack explicit alignment with linguistic units. To enhance the flexibility of speculative decoding for speech synthesis without compromising output quality, we introduce a tolerance factor into the criteria.



Given an input prompt \(T_{in}\) in concatenated by text and speech tokens, the draft model first generates \(L_d\) candidate tokens autoregressively:
\begin{align}
    p^d_1, \ldots, p^d_{L_d} &= M_{p}(T_{in}), \\
    x_1, \ldots, x_{L_d} &\sim p^d_1, \ldots, p^d_{L_d}.
\end{align}
where \(x_i\) is a draft token, \(p^d_i\) is the probability distribution of the draft model, and \(i \in [1, L_d]\).

After we obtain the draft tokens, the target model \(M_{q}\) then processes all draft tokens in parallel to produce verification distributions \(q^s\):
\begin{equation}
    q^s_1, \ldots, q^s_{L_d + 1} = M_{q}(T_{in}, x_1, \ldots, x_{L_d}).
\end{equation}

Unlike text generation where tokens have discrete semantics and one-to-one mappings with words or phonemes, speech tokens exhibit continuous and overlapping characteristics. Strict acceptance criteria in standard speculative decoding often lead to unnecessary rejections, however minor distribution discrepancies may not perceptually affect speech quality. Therefore, we introduce a tolerance factor $\beta \geq 0$ that relaxes the acceptance threshold:
\begin{equation}
\text{Accept } x_i \text{ if } r_i < \min(1, q^s_i / p^d_i) + \beta,
\end{equation}
where \(r_i \sim U[0,1]\). This modification allows acoustically plausible variations to be accepted, increasing decoding throughput without compromising perceptual quality.


When rejection occurs at position \(i\), we resample token \(x'_i\) from the adjusted distribution:
\begin{equation}
    \text{Resample } x'_i \sim 
    p_{new} = \text{normlize}(\max(0, q_i^s - p_i^d)),
\end{equation}
ensuring the corrected token follows the target model's distribution. If all \(L_d\) tokens are accepted, we append an additional token sampled from \(q^s_{L_d + 1}\).

\subsection{Lightweight Draft Model Construction}

Modern autoregressive speech synthesis systems rely on a single large target model, posing a challenge for speculative decoding: constructing a lightweight draft model that aligns with the vocabulary of target model. To address this, we design the draft model as a shallower variant of the target architecture, sharing its core structure but with fewer Transformer layers. Crucially, the draft model is initialized using the pre-trained upper layers of the target model (e.g., layers 19–24 in a 24-layer setup), enabling immediate compatibility with its linguistic representations while avoiding cold-start training.

To enhance coordination between the models without costly full fine-tuning, we employ a layer-specific adaptation strategy inspired by adapter tuning \cite{houlsby2019parameter,thomas2022efficient}. During training, only lower layers (e.g., layers 1–2) of the draft model and classification head are updated, while the initialized upper layers and embeddings remain frozen. This approach preserves high-level semantic knowledge of the target model while allowing the draft model to adapt to speech-specific patterns through tunable lower layers. Unlike traditional adapter methods that insert new modules, our strategy achieves parameter efficiency through selective layer inheritance and hierarchical fine-tuning. To keep the training stage simple and efficient, we only use the cross entropy loss to fine-tune the draft model: 
\begin{equation}
    L_{draft} = -\tfrac{1}{L_{tgt}}\textstyle\sum_{l=1}^{L_{tgt}}\log(M_p(u_{<l}))
\end{equation}
where $U_{tgt} = \{u_1, ..., u_{L_{tgt}}\}$ is the target sequence, $u_{<l} = \{u_1, ..., u_{l-1}\}$, and $L_{tgt}$ is the length of the target sequence.


\section{Experiments}


\subsection{Implementation Details}

We use CosyVoice 2 \cite{du2024cosyvoice2} as the target model, which is a highly efficient TTS model consisting of 24 layers of Transformer adapted from Qwen2.5 (0.5B) \cite{yang2024qwen2}. Based on this model, we design a draft model consisting of 8 layers of Transformer. We follow the setting of CosyVoice 2: each Transformer layer typically employs a multi-head attention mechanism with the number of heads set to 14, and the feed-forward layer in each Transformer layer has a hidden dimension of 896. We train the draft model on LibriTTS \cite{zen2019libritts} while keeping the parameters of CosyVoice 2 frozen. To accelerate the training process, we load the pre-trained parameters of CosyVoice 2 to initialize the model. For the Transformer layers, we utilize the parameters of the lower two layers and the upper six layers of CosyVoice 2. During the training of the draft model, we only train the first two layers of the Transformer and the classification layer. The length of draft tokens as $L_d$ is set to 3 and the tolerance factor $\beta$ is set to 0.4. We conduct all the experiments on NVIDIA A100-SXM4-40GB GPU.

\subsection{Dataset}

LibriTTS, introduced by \cite{zen2019libritts}, is a well-known multi-speaker English speech dataset with transcriptions. It encompasses a substantial amount of speech, sourced from 2,456 English speakers, 585 hours in total.

For data partitioning, the subsets “train-clean-100”, “train-clean-360”, and “train-other-500” are grouped together to form the training data. This combined training set is approximately 580 hours long and is contributed by 2,306 speakers. And we merge the subsets “dev-clean” and “dev-other” for model selection, based on token prediction accuracy of the speech token sequences on this set.

Following \cite{du2024unicats}, we evaluate SSD on the subset of LibriTTS. This testset, which covers 37 speakers, contains 500 utterances selected from the “test-clean” subset of LibriTTS. Its purpose is to evaluate the zero-shot adaptation ability for entirely new and unseen speakers. Moreover, for each speaker in the testset, there is a short voice prompt lasting approximately 3 seconds.

\subsection{Evaluation Metrics}

Referring to \cite{du2024cosyvoice2}, we take word error rate (WER), speaker similarity (SS), and NMOS score as the objective metrics for a comprehensive evaluation of SSD. The Word Error Rate can measure content consistency. We use Whisper-large V3 as the Automatic Speech Recognition (ASR) model to calculate the WER. Regarding the Speaker Similarity, we employ the ERes2Net model to extract the speaker embedding vectors of the prompt speech and the generated one, and adopt their raw cosine similarity as the speaker similarity. The NMOS score is a widely-adopted metric for evaluating the objective quality of speech. We compute LM-RTF to assess the efficiency of the LM, which is defined as the ratio of the time taken by the LM to generate the speech to the duration of the synthesized speech.

For subjective evaluation, we conduct two Mean Opinion Score (MOS) tests on a 5-point scale with 1-point intervals (1: lowest, 5: highest), involving 30 raters each evaluating 10 randomized samples. First, we focus on the similarity (SIM-MOS), which provides insights into how closely the synthetic voice mimics the unique vocal characteristics of the target speaker. Then, we evaluate naturalness of the speech (NAT-MOS), meaning how close the synthesized voice is to a real-human voice and detecting any signs of artificial synthesis.

\vspace{-5pt}
\section{Results and Analysis}

We apply SSD to CosyVoice 2 and compare the result with draft model and CosyVoice 2 using autoregressive decoding to synthesize speech.

\subsection{Objective Evaluation}


As shown in \autoref{tab:Objective}, our Speech Speculative Decoding (SSD) framework achieves 1.4× faster synthesis than the conventional autoregressive decoding used by CosyVoice 2. Notably, this acceleration is attained despite SSD being trained on a small-scale open-source dataset (approximately 1/300 the size of CosyVoice 2’s proprietary industrial dataset \cite{du2024cosyvoice2}), highlighting its robust compatibility with large-scale systems. While the WER of our system shows a slight increase compared to CosyVoice 2, the negligible differences in NMOS and SS, coupled with the performance in SIM-MOS and NAT-MOS as discussed in Section \ref{subsec:subeval}, suggest that the WER discrepancy stems primarily from extrapolation generalization limitations due to limited training data.


Critically, despite that the draft model achieves the worst WER (16.13\%), SSD maintains robust synthesis quality through its verification mechanism. This decoupling between draft model performance and final output quality underscores robustness of SSD: even with a suboptimal draft model, the framework maintains strong synthesis quality while delivering significant acceleration. These results collectively validate the effectiveness of SSD as a promising solution for autoregressive speech synthesis acceleration.

\begin{table}[h]
\centering
\vspace{-6pt}
\caption{Objective evaluation result of SSD with $\beta = 0.4$.}
\vspace{-6pt}
\footnotesize
\begin{tabular}{c|ccc|c}
\hline
\textbf{Model} & \textbf{WER (\%)}$\downarrow$ & \textbf{NMOS} $\uparrow$ & \textbf{SS} $\uparrow$ & \textbf{LM-RTF} $\downarrow$ \\
\hline
Draft Model & 16.13 & 3.95 & 0.61 & 0.222 \\
CosyVoice 2  & \textbf{3.67} & \textbf{3.96} & 0.62 & 0.504 \\
\quad w. SSD  & 5.70 & 3.94 & \textbf{0.63} & \textbf{0.360} \\
\hline
\end{tabular}
\label{tab:Objective}
\vspace{-17pt}
\end{table}

\subsection{Subjective Evaluation}
\label{subsec:subeval}

Results in
\autoref{tab:Subjective} demonstrate that SSD achieves comparable performance to the original CosyVoice 2 using autoregressive decoding in terms of both similarity (SIM-MOS) and naturalness (NAT-MOS). While the draft model alone achieves the lowest scores, highlighting its limitations in independent high-quality speech generation, SSD bridges this gap by leveraging the target model for verification. These results confirm that SSD effectively preserves the synthesis quality of the target model while accelerating inference, validating it as a practical solution for efficient autoregressive speech synthesis without compromising perceptual quality.

\begin{table}[htbp]
\centering
\footnotesize
\vspace{-6pt}
\caption{Subjective evaluation result of SSD with $\beta = 0.4$.}
\vspace{-6pt}
\begin{tabular}{c|cc}
\hline
\textbf{Model} & \textbf{SIM-MOS} $\uparrow$ & \textbf{NAT-MOS }$\uparrow$ \\
\hline
Draft Model  & 3.737 $\pm$ 0.141 & 3.820 $\pm$ 0.143 \\
CosyVoice 2  & \textbf{3.789} $\pm$ 0.148 & \textbf{3.930} $\pm$ 0.147 \\
\quad w. SSD & 3.784 $\pm$ 0.152 & 3.925 $\pm$ 0.139 \\
\hline
\end{tabular}
\label{tab:Subjective}
\vspace{-18pt}
\end{table}

\subsection{Analysis of Tolerance Factor}

We further analyze how tolerance factors influence model acceleration and performance in this section. \autoref{tab:tf_obj} shows the objective evaluation result with different tolerance factors $\beta$. We conduct this experiment with the length of draft tokens $L_d$ set to 3. For the efficiency of the model, as the tolerance factor increases, more words are accepted during the speculative decoding process, and the efficiency of the model in synthesizing speech gradually improves.

\begin{table}[htbp] 
    \centering 
    \footnotesize
    \caption{Objective evaluation result of SSD with different tolerance factor $\beta$.} 
    \label{tab:tf_obj} 
    \vspace{-7pt}
    \begin{tabular}{c|ccc|c}
        \toprule 
        $\boldsymbol{\beta}$ & \textbf{WER(\%)}$\downarrow$ & \textbf{NMOS}$\uparrow$ & \textbf{SS}$\uparrow$ & \textbf{LM-RTF}$\downarrow$\\
        \midrule 
        $\beta=0.0$ & 6.34 & 3.86 & 0.62 & 0.509\\
        $\beta=0.1$ & 5.71 & 3.90 & 0.50 & 0.444\\
        $\beta=0.2$ & 6.08 & 3.92 & 0.62 & 0.408\\
        $\beta=0.3$ & 6.29 & 3.93 & 0.62 & 0.386\\
        $\beta=0.4$ & \textbf{5.70} & \textbf{3.94} & \textbf{0.63} & \textbf{0.360}\\
        \bottomrule
    \end{tabular}
    \vspace{-10pt}
\end{table}

Notably, experimental results indicate that varying tolerance factors do not lead to significant differences in speech quality, naturalness, or speaker similarity. Specifically, as the tolerance factor increases, more tokens generated by the draft model are accepted, leading to divergent token distributions across different $\beta$. However, the speech quality across different $\beta$ remains consistent, suggesting that a single speech token does not exhibit a one-to-one correspondence with a word or a specific type of paralinguistic information. Instead, multiple combinations of speech tokens can reconstruct similar speech signals through the speech decoder. Thus, this characteristic of the autoregressive speech synthesis system makes SSD a highly rational and promising approach for accelerating the inference process of autoregressive speech synthesis models, as SSD allows for more diverse token generation.

\subsection{Analysis of the Length of Draft Tokens}

\begin{figure}[htbp]
  \vspace{-5pt}
  \centering
  \includegraphics[width=0.8\linewidth]{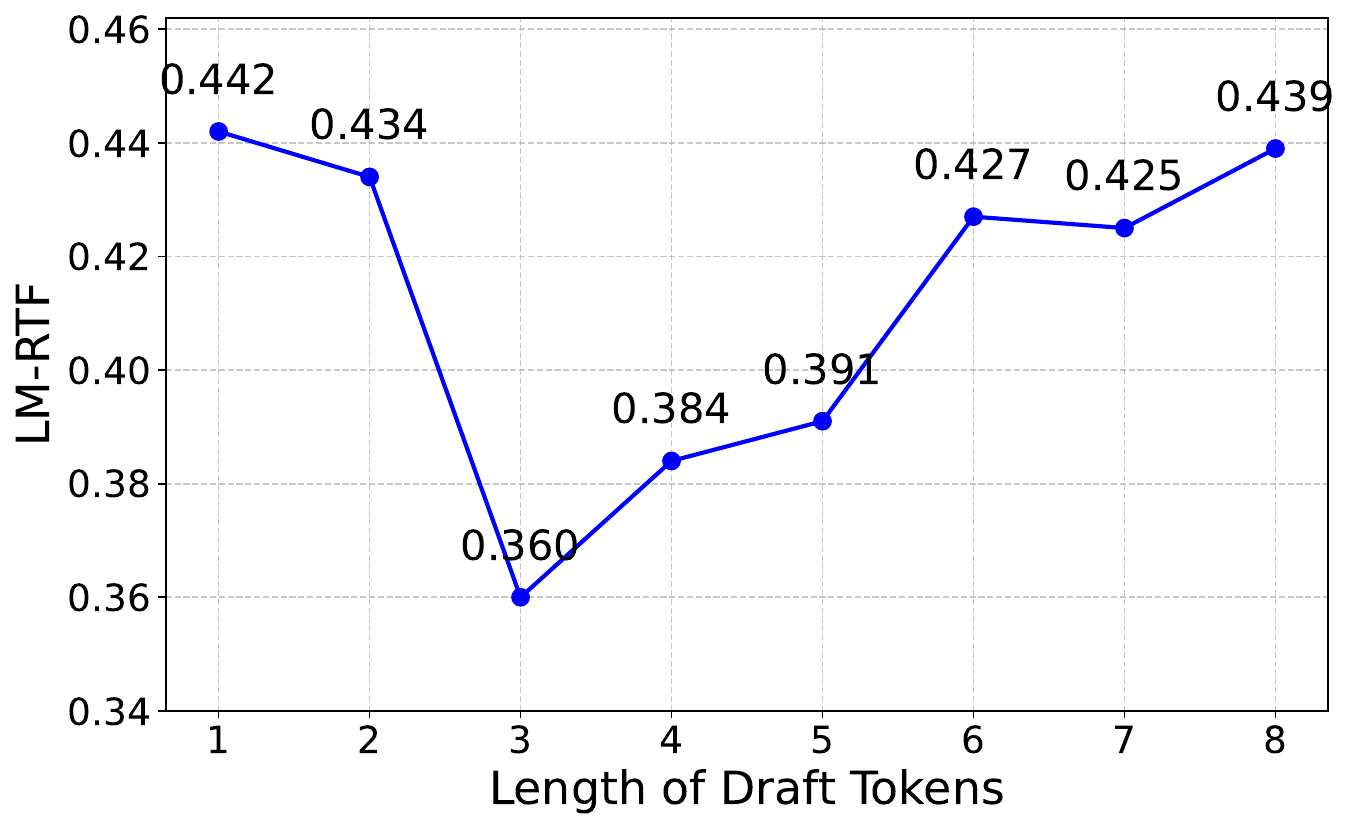}
  \vspace{-10pt} 
  \caption{LM-RTF results of SSD with different $L_d$.}
  \label{fig:draft_len_lmrtf}
  \vspace{-15pt}
\end{figure}

To explore the trade-off between draft token length and model efficiency, we analyze acceleration results with varying $L_d$. As shown in \autoref{fig:draft_len_lmrtf}, the efficiency of the model initially ascends and then declines. This is attributed to the fact that as $L_d$ grows larger, the quality of the draft tokens decrease, leading to the generation of many useless tokens due to error accumulation. On the other hand, shorter draft lengths result in higher token acceptance rates but fail to fully leverage the parallel verification capability for accelerating inference. Thus, draft length represents a trade-off between draft model quality and verification model acceleration.


\section{Conclusions}

In this work, we introduce Speculative Speech Decoding (SSD), a novel approach to accelerate autoregressive speech synthesis. We propose a lightweight draft model constructed by fine-tuning a few parameters from the target model. Experiments on CosyVoice 2 show that our proposed SSD achieves 1.4x acceleration while maintaining high synthesis quality. Subjective evaluations demonstrate that the speech generated by SSD achieves comparable performance to the original CosyVoice 2 in terms of similarity and naturalness. SSD demonstrates a reliable and promising solution for more efficient speech synthesis.

\section{Acknowledgements}
This work is supported by National Natural Science Foundation of China (62076144), National Social Science Foundation of China (13\&ZD189) and Shenzhen Science and Technology Program (JCYJ20220818101014030).

\bibliographystyle{IEEEtran}
\bibliography{mybib}

\end{document}